\documentclass{elsart3p}

\usepackage{amssymb}
\usepackage[dvips]{hyperref}
\usepackage{hypernat}
\usepackage[dvips]{graphicx}
\usepackage[T1]{fontenc}
\begin{document}

\begin{frontmatter}


 \title{Exchange bias and asymmetric hysteresis loops from a microscopic model of core/shell nanoparticles}
\author[1]{\`Oscar Iglesias},
\ead{oscar@ffn.ub.es}
\ead[url]{http://www.ffn.ub.es/oscar}
\corauth[1]{Corresponding author: Tel. +34-934039201}
\author{Xavier Batlle and Am\'{\i}lcar Labarta}
\address{Departament de F\'{\i}sica Fonamental and Institut de Nanoci\`encia i Nanotecnologia (IN2UB), Universitat de Barcelona, Av. Diagonal 647, 08028 Barcelona (Spain)}


\begin{abstract}
We present Monte Carlo simulations of hysteresis loops of a model of a magnetic nanoparticle with a ferromagnetic core and an antiferromegnetic shell with varying values of the core/shell interface exchange coupling which aim to clarify the microscopic origin of exchange bias observed experimentally. We have found loops shifts in the field direction as well as displacements along the magnetization axis that increase in magnitude when increasing the interfacial exchange coupling. Ovelap functions computed from the spin configurations along the loops have been computed to explain the origin and magnitude of these features microscopically.

\end{abstract}

\end{frontmatter}

\section{Introduction}

Compound magnetic structures formed by a ferromagnet (FM) in close contact with an antiferromagnet (AFM) show interesting proximity effects which are enhanced when going to the nanoscale range \cite{Nogues_physrep05}. Among them, one of the most studied phenomenon is the displacement of the hysteresis loops along the field axis observed in layered FM/AFM systems and core/shell nanoparticles when measured after cooling in a magnetic field, an effect that has been termed exchange bias (EB) after its first observation by Meiklejohn and Bean in Co particles \cite{Meiklejohn}. In spite of the intensive research developed in the last decades, there are still many open issues to be understood and no clear-cut model has been able to account for all the observed phenomenology \cite{KiwiStamps}.
Several recent experiments based on spectrocopic techniques  \cite{Ohldag} have focused on the observation of the magnetization reversal at the FM/AFM interface in different layered systems since it is widely accepted that the magnitude of the exchange bias field is related to the existence of uncompensated moments at the interface. 
However, in nanoparticles with FM core/AFM shell structure, details about the core/shell interface cannot be accessed by these techniques and only indirect information about the magnetic order at the inerface can be gained by magnetization measurements. 

In this study, we present the results of Monte Carlo simulations of hysteresis loops of an individual core/shell nanoparticles, which, apart from the usual shift in the field axis direction, display asymmetry between the positive and negative field branches and also shifts in the vertical direction as observed in some experimental systems \cite{Zheng}. By inspection of the spin configurations during field cycling, we will show how this phenomenolgy is related to the peculiar interplay between exchange coupling at the core/shell interface and the magnetic order of its uncompensated spins induced after the field cooling process.

\section{Model and Results}
The details of model for the core/shell particle on which simulations are based were already presented in our previous studies \cite{Iglesias}, where the details about the simulation protocol used for the Monte Carlo method can also be found.
The simulations are based on the following Hamiltonian:
\begin{eqnarray*}
\frac{H}{k_B}= -\sum_{\langle  i,j\rangle}J_{ij}\ {\vec S}_i \cdot {\vec S}_j
-\sum_{i}k_i\ (S_i^z)^2 -\sum_{i}\vec h\cdot{\vec S_i}\ ,
\end{eqnarray*}
where ${\vec S}_i$ are classical Heisenberg spins of unit magnitude placed at the nodes of a sc lattice. In the first term, the value of the n.n. exchange constants $J_{ij}$ depends on the spins belonging to different particle regions. At the core, $J_{ij}$ is FM and has been fixed to $J_{\mathrm{C}}= +10$ K. AF spins at the shell have $J_{\rm{Sh}}= -0.5 J_{\mathrm{C}}$. The exchange coupling at the interface [defined as those spinson the core (shell) with at least one neighbor on the shell(core)] is $J_{\mathrm{Int}}$. 

\begin{figure}[t]
\includegraphics[width=\columnwidth]{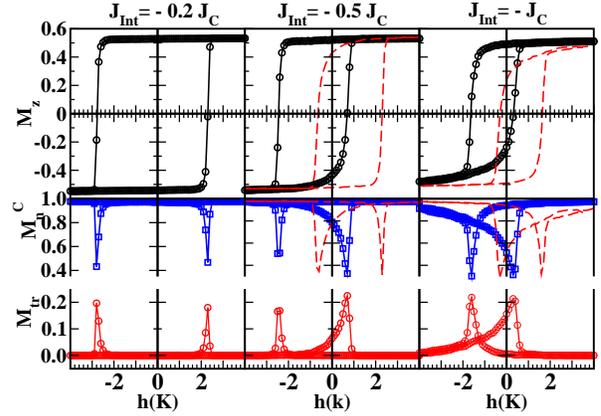}
\caption{\label{Fig_1}(Color online) Upper panels display the hysteresis loops for a particle with radius $R=12\,a$ obtained after FC down to $T= 0.05$ in a field $h_\mathrm{FC}= 4$ (circles) and $h_\mathrm{FC}= -4$ (dashed lines) for three values of the exchange coupling constant $J_\mathrm{Int}$ at the core/shell interface. Lower panels show the average magnetization projection of the core spins along the field axis $m_\mathrm{n}^C$ (squares) and the hysteresis loops for the  component of the magnetization transverse to the field direction $M_\mathrm{tr}$ (circles).}
\end{figure}

Hysteresis loops obtained for a particle with total radius $R= 12$a, surface shell thickness $R_\mathrm{Sh}=3$a and $J_{\mathrm{Sh}}= -0.5 J_{\mathrm{C}}$, obtained after cooling from above the Ne\'el temperature of the AF shell in a field $h_\mathrm{FC}= 4$ are shown in Fig. \ref{Fig_1} (circles)for three values of $J_{\mathrm{Int}}$. 
As can be clearly seen in the upper panels of the figure, the loops display an increasing shift towards the negative field direction with increasing values of $J_{\mathrm{Int}}$, demonstrating that the observance of EB is related to the AF coupling of the shell spins the core spins at the interface. 

A detailed inspection of the microscopic configurations attained after the FC process \cite{Iglesias} revealed that, inspite of the AF character of the coupling between the shell spins and also between the interfacial spins, a net magnetization component along the field direction is induced at the shell interface due to the geometrical symmetry breaking and the alignment of groups of spins into the field direction. This net magnetic moment generates local fields on the core spins that point into the same direction as the external field, causing the shift of the hysteresis loops. This is further supported by the observation that the hysteresis loops are shifted by the same amount but towards the positive field axis when obtained after cooling in a field applied in a direction negative with respect to the measuring field (see for example the dashed lines in Fig \ref{Fig_1}. for $J_{\mathrm{Int}}/J_{\mathrm{C}}= -0.5, -1$).

\begin{figure}[tbp]
\includegraphics[width=0.9\columnwidth]{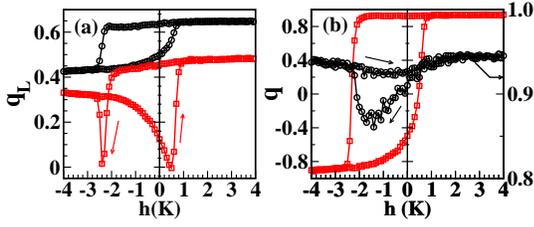}
\caption{\label{Fig_2}(Color online) Field dependence of the link overlap $q_\mathrm L$ (a) and overlap $q$ (b) functions for the interfacial spins at the shell (circles) and at the core (squares) for $J_\mathrm{Int}= -0.5 J_\mathrm{C}$.
}
\end{figure}

In addition, a clear asymmetry between the upper and lower loop branches developes with increasing values of the interface coupling. This feature can be more clearly seen by looking at the average absolute value of the magnetization projection along the field axis through the reversal process, $M_n^{\mathrm C}=\sum_i |\vec{S_i}\cdot\hat{z}|$ displayed in the lower panels of Fig. \ref{Fig_1} for the core spins. 
The origin of the loop asymmetry can be traced by monitoring the values the so-called overlap $q(h)$ and link overlap $q_L(h)$ functions along the hysteresis loops, that are a generalization of similar quantities commonly used in the spin-glass literature \cite{Katzgraber} and that are defined as  
$q(h) = \frac{1}{N} \sum_{i=1}^N \vec{S_i}(h_\mathrm{FC})\cdot \vec{S_i}(h)$ and
$q_L(h) =  \sum_{\langle ij \rangle} \frac{1}{N_l}\ \vec{S_i}(h_\mathrm{FC})\cdot \vec{S_j}(h_\mathrm{FC})\ \vec{S_i}(h)\cdot\vec{S_j}(h)$,	
where in $q_L(h)$ the summation is over nearest neighbors and $N_l$ is a normalization factor that counts the number of bonds. An example of the field dependence of these overlaps computed only for the interfacial spins is shown in Fig. \ref{Fig_2}, where we have separated the contribution of the shell and core spins.
A departure of $q_L$ from unity is known to be proportional to the surface of reversed domains formed at field $h$ and, therefore, $q_L$ is sensible to the existence of domain walls.
The sharp decrease of $q_L$ for core spins and the symmetry of the peak around the negative coercive field indicates uniform reversal. However, the progressive reduction of $q_L$ along the ascending branch and the asymmetry of the peak around the positive coercive field indicates the formation of domain walls at the particle core that sweep the particle during reversal.
The function $q_(h)$ measures differences of the spin configuration at field $h$ with respect to the one attained after FC. Therefore, the decrease of $q$ for the interface shell spins when reducing the magnetic field indicates the existence of a fraction of shell spins that reverse dragged by core spins, while the constancy of $q$ in the ascending branch reveals the existence of spins pinned during core reversal. 

\begin{figure}[tbp]
\includegraphics[height=\columnwidth,angle= -90]{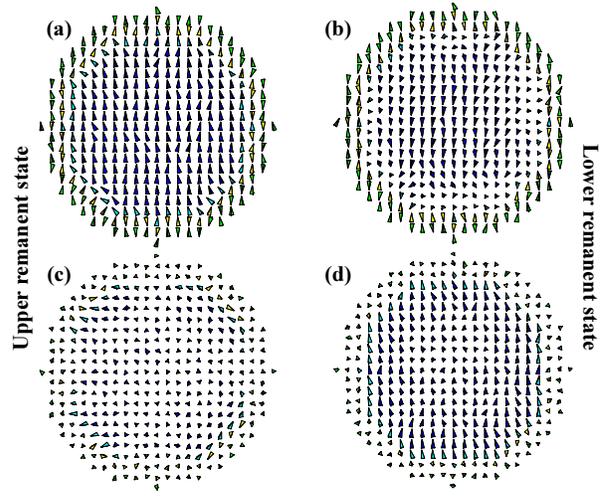}
\caption{\label{Fig_3}(Color online) Snapshots of the remanent spin configurations of the upper (a,c panels) and lower (b,d panels) branches of the hystersis loops showing midplane cross sections of the particle parallel (a,b) and perpendicular to the z axis (c, d) for the case $J_{\mathrm {Int}}= - J_\mathrm{C}$ shown in Fig. 1a. Dark (light) blue cones represent core (core/interface) spins while green (yellow) ones are for spins at the shell (shell/interface).
}
\end{figure}

Clearly correlated to the observation of EB and the loop asymmetry, the loops experience a shift along the vertical ($M_z$) axis which increases with $J_\mathrm{Int}$, as reflected in Fig. \ref{Fig_1} by the difference of the $M_z$ values in the high field region of the two loop branches or at the remanence points. 
The field dependence of magnetization transverse to the field direction $M_\mathrm{tr}$ (circles in the lower panels of Fig. \ref{Fig_1}), indicates that $M_\mathrm{tr}$ has higher values for the descending loop branch that in the ascending branch, and that $M_\mathrm{tr}$ increases with $J_\mathrm{Int}$. 
Snapshots of the spin configurations at the remanence points shown in Fig. \ref{Fig_3}, show the existence of a higher amount of core spins with transverse orientation near the interface at the lower branch (Fig. \ref{Fig_3}b, d) than at the upper branch. This is in agreementwith the results of some experiments of oxidized particles where this vertical shift was also observed \cite{Zheng}. Our simulation results above, indicate that the microscopic origin of the vertical shift is the due to the different reversal mechanisms on the two loop branches due to the existence of uncompensated pinned moments at the core/shell interface that faciltate the nucleation of non-uniform magnetic structures during the ascending field branch of the loops.

We acknowledge H. Katzgraber for helpful comments on overlap functions and CESCA and CEPBA under coordination of C$^4$ for computer facilities. This work has been supported by the Spanish MEyC through the MAT2003-01124 project and the Generalitat de Catalunya through the 2005SGR00969 CIRIT project.


\end{document}